\documentclass[%
    reprint,
     amsmath,amssymb,
     aps,
    floatfix
]{revtex4-2}

\usepackage{graphicx}% Include figure files
\usepackage{dcolumn}% Align table columns on decimal point
\usepackage{bm}% bold math
\usepackage[]{units}
\usepackage{xcolor}
\usepackage{tabulary}
\usepackage{comment}
\usepackage[T5,T1]{fontenc}

\begin{document}

\title{Cosmic Ray Spectrum from 250 TeV to 10 PeV using IceTop}

\affiliation{III. Physikalisches Institut, RWTH Aachen University, D-52056 Aachen, Germany}
\affiliation{Department of Physics, University of Adelaide, Adelaide, 5005, Australia}
\affiliation{Dept. of Physics and Astronomy, University of Alaska Anchorage, 3211 Providence Dr., Anchorage, AK 99508, USA}
\affiliation{Dept. of Physics, University of Texas at Arlington, 502 Yates St., Science Hall Rm 108, Box 19059, Arlington, TX 76019, USA}
\affiliation{CTSPS, Clark-Atlanta University, Atlanta, GA 30314, USA}
\affiliation{School of Physics and Center for Relativistic Astrophysics, Georgia Institute of Technology, Atlanta, GA 30332, USA}
\affiliation{Dept. of Physics, Southern University, Baton Rouge, LA 70813, USA}
\affiliation{Dept. of Physics, University of California, Berkeley, CA 94720, USA}
\affiliation{Lawrence Berkeley National Laboratory, Berkeley, CA 94720, USA}
\affiliation{Institut f{\"u}r Physik, Humboldt-Universit{\"a}t zu Berlin, D-12489 Berlin, Germany}
\affiliation{Fakult{\"a}t f{\"u}r Physik {\&} Astronomie, Ruhr-Universit{\"a}t Bochum, D-44780 Bochum, Germany}
\affiliation{Universit{\'e} Libre de Bruxelles, Science Faculty CP230, B-1050 Brussels, Belgium}
\affiliation{Vrije Universiteit Brussel (VUB), Dienst ELEM, B-1050 Brussels, Belgium}
\affiliation{Dept. of Physics, Massachusetts Institute of Technology, Cambridge, MA 02139, USA}
\affiliation{Dept. of Physics and Institute for Global Prominent Research, Chiba University, Chiba 263-8522, Japan}
\affiliation{Department of Physics, Loyola University Chicago, Chicago, IL 60660, USA}
\affiliation{Dept. of Physics and Astronomy, University of Canterbury, Private Bag 4800, Christchurch, New Zealand}
\affiliation{Dept. of Physics, University of Maryland, College Park, MD 20742, USA}
\affiliation{Dept. of Astronomy, Ohio State University, Columbus, OH 43210, USA}
\affiliation{Dept. of Physics and Center for Cosmology and Astro-Particle Physics, Ohio State University, Columbus, OH 43210, USA}
\affiliation{Niels Bohr Institute, University of Copenhagen, DK-2100 Copenhagen, Denmark}
\affiliation{Dept. of Physics, TU Dortmund University, D-44221 Dortmund, Germany}
\affiliation{Dept. of Physics and Astronomy, Michigan State University, East Lansing, MI 48824, USA}
\affiliation{Dept. of Physics, University of Alberta, Edmonton, Alberta, Canada T6G 2E1}
\affiliation{Erlangen Centre for Astroparticle Physics, Friedrich-Alexander-Universit{\"a}t Erlangen-N{\"u}rnberg, D-91058 Erlangen, Germany}
\affiliation{Physik-department, Technische Universit{\"a}t M{\"u}nchen, D-85748 Garching, Germany}
\affiliation{D{\'e}partement de physique nucl{\'e}aire et corpusculaire, Universit{\'e} de Gen{\`e}ve, CH-1211 Gen{\`e}ve, Switzerland}
\affiliation{Dept. of Physics and Astronomy, University of Gent, B-9000 Gent, Belgium}
\affiliation{Dept. of Physics and Astronomy, University of California, Irvine, CA 92697, USA}
\affiliation{Karlsruhe Institute of Technology, Institut f{\"u}r Kernphysik, D-76021 Karlsruhe, Germany}
\affiliation{Dept. of Physics and Astronomy, University of Kansas, Lawrence, KS 66045, USA}
\affiliation{SNOLAB, 1039 Regional Road 24, Creighton Mine 9, Lively, ON, Canada P3Y 1N2}
\affiliation{Department of Physics and Astronomy, UCLA, Los Angeles, CA 90095, USA}
\affiliation{Department of Physics, Mercer University, Macon, GA 31207-0001, USA}
\affiliation{Dept. of Astronomy, University of Wisconsin, Madison, WI 53706, USA}
\affiliation{Dept. of Physics and Wisconsin IceCube Particle Astrophysics Center, University of Wisconsin, Madison, WI 53706, USA}
\affiliation{Institute of Physics, University of Mainz, Staudinger Weg 7, D-55099 Mainz, Germany}
\affiliation{Department of Physics, Marquette University, Milwaukee, WI, 53201, USA}
\affiliation{Institut f{\"u}r Kernphysik, Westf{\"a}lische Wilhelms-Universit{\"a}t M{\"u}nster, D-48149 M{\"u}nster, Germany}
\affiliation{Bartol Research Institute and Dept. of Physics and Astronomy, University of Delaware, Newark, DE 19716, USA}
\affiliation{Dept. of Physics, Yale University, New Haven, CT 06520, USA}
\affiliation{Dept. of Physics, University of Oxford, Parks Road, Oxford OX1 3PU, UK}
\affiliation{Dept. of Physics, Drexel University, 3141 Chestnut Street, Philadelphia, PA 19104, USA}
\affiliation{Physics Department, South Dakota School of Mines and Technology, Rapid City, SD 57701, USA}
\affiliation{Dept. of Physics, University of Wisconsin, River Falls, WI 54022, USA}
\affiliation{Dept. of Physics and Astronomy, University of Rochester, Rochester, NY 14627, USA}
\affiliation{Oskar Klein Centre and Dept. of Physics, Stockholm University, SE-10691 Stockholm, Sweden}
\affiliation{Dept. of Physics and Astronomy, Stony Brook University, Stony Brook, NY 11794-3800, USA}
\affiliation{Dept. of Physics, Sungkyunkwan University, Suwon 16419, Korea}
\affiliation{Institute of Basic Science, Sungkyunkwan University, Suwon 16419, Korea}
\affiliation{Dept. of Physics and Astronomy, University of Alabama, Tuscaloosa, AL 35487, USA}
\affiliation{Dept. of Astronomy and Astrophysics, Pennsylvania State University, University Park, PA 16802, USA}
\affiliation{Dept. of Physics, Pennsylvania State University, University Park, PA 16802, USA}
\affiliation{Dept. of Physics and Astronomy, Uppsala University, Box 516, S-75120 Uppsala, Sweden}
\affiliation{Dept. of Physics, University of Wuppertal, D-42119 Wuppertal, Germany}
\affiliation{DESY, D-15738 Zeuthen, Germany}

\author{M. G. Aartsen}
\affiliation{Dept. of Physics and Astronomy, University of Canterbury, Private Bag 4800, Christchurch, New Zealand}
\author{R. Abbasi}
\affiliation{Department of Physics, Loyola University Chicago, Chicago, IL 60660, USA}
\author{M. Ackermann}
\affiliation{DESY, D-15738 Zeuthen, Germany}
\author{J. Adams}
\affiliation{Dept. of Physics and Astronomy, University of Canterbury, Private Bag 4800, Christchurch, New Zealand}
\author{J. A. Aguilar}
\affiliation{Universit{\'e} Libre de Bruxelles, Science Faculty CP230, B-1050 Brussels, Belgium}
\author{M. Ahlers}
\affiliation{Niels Bohr Institute, University of Copenhagen, DK-2100 Copenhagen, Denmark}
\author{M. Ahrens}
\affiliation{Oskar Klein Centre and Dept. of Physics, Stockholm University, SE-10691 Stockholm, Sweden}
\author{C. Alispach}
\affiliation{D{\'e}partement de physique nucl{\'e}aire et corpusculaire, Universit{\'e} de Gen{\`e}ve, CH-1211 Gen{\`e}ve, Switzerland}
\author{N. M. Amin}
\affiliation{Bartol Research Institute and Dept. of Physics and Astronomy, University of Delaware, Newark, DE 19716, USA}
\author{K. Andeen}
\affiliation{Department of Physics, Marquette University, Milwaukee, WI, 53201, USA}
% ...
\date{\today}

\collaboration{IceCube Collaboration}
\noaffiliation

\begin{abstract}
	
	We report here an extension of the measurement of the all-particle cosmic-ray spectrum with IceTop to lower energy. The new measurement gives full coverage of the knee region of the spectrum and reduces the gap in energy between previous IceTop and direct measurements.  With a new trigger that selects events in closely spaced detectors in the center of the array, the IceTop energy threshold is lowered by almost an order of magnitude below its previous threshold of 2~PeV. In this paper we explain how the new trigger is implemented, and we describe the new machine-learning method developed to deal with events with very few detectors hit. We compare the results with previous measurements by IceTop and others that overlap at higher energy and with HAWC and Tibet in the 100 TeV range.
	
\end{abstract}

\maketitle

\section{Introduction}\label{section_intro}
    Cosmic rays are charged particles that reach Earth from space with energies as high as hundreds of EeV. The sources of high energy cosmic rays and their acceleration mechanism are not fully known, but they are reflected in the all-particle cosmic ray energy spectrum measured by ground-based air shower experiments. The differential energy spectrum is characterized as a power law, $\frac{{\rm d}N}{{\rm d}E} \propto E^{-\gamma}$, where $\gamma$ is the differential spectral index. Features in the spectrum correspond to changes in $\gamma$. Around \unit[3$\times10^{15}$]{eV}, $\gamma$ increases from $\sim$2.7 to $\sim$3.0 and creates a knee-like structure, first mentioned in 1958 by Kulikov and Khristiansen~\cite{knee_first_observed}. Similarly, around 5$\times$\unit[10$^{18}$]{eV}, $\gamma$ decreases from $\sim$3.0 and creates an ankle-like structure~\cite{Abbasi:2007sv, Abraham:2010mj}. This analysis covers the energy region around the knee.
    
    The transition from galactic to extragalactic sources happens somewhere between the knee and the ankle, but the exact nature of the transition is unknown. The ankle is believed to be the energy region above which cosmic rays are mostly coming from extra-galactic sources~\cite{HILLAS2004139}. Since propagation and acceleration both depend on magnetic fields, the spectra of individual elements are expected to depend on magnetic rigidity~\cite{Peters1961}. 
    
    The cosmic ray energy spectrum and its chemical composition are measured directly up to \unit[100]{TeV} using detectors in satellites and balloons.  The flux decreases sharply as energy increases, so indirect measurements with large arrays of detectors on the ground are needed for higher energies. There are several ground-based cosmic ray detectors sensitive to cosmic rays above a few TeV. For example, HAWC~\cite{HAWCDetector_1, HAWCDetector_2} is a ground-based gamma ray and cosmic ray detector that measures cosmic rays from \unit[10]{TeV} to \unit[500]{TeV}~\cite{HAWC2017}. The threshold of Tibet~III~\cite{Amenomori} is \unit[100]{TeV}, and its measurement extends through the knee region.  
    KASCADE~\cite{KASCADE2003} and KASCADE-Grande~\cite{Apel:2012tda} measure the energy spectrum in the range of hundreds of TeV to EeV~\cite{kascade_spectrum,Apel:2012tda}. The Telescope Array~\cite{ta_detector} and the Pierre Auger Observatory~\cite{auger_detector} detect ultra high energy cosmic rays with energy higher than \unit[100]{PeV}~\cite{TA2017, Auger2012}. 
	The low-energy extension of TA (TALE)~\cite{tale_spectrum} connects these ultra-high energy measurements with the knee region. The combined energy spectra from all detectors provides an overview of the origin and the acceleration mechanism of cosmic rays.  
    
    The IceTop energy spectrum thus far covers an energy region from \unit[2]{PeV} to \unit[1]{EeV}~\cite{bhaktiyar, IceCube_3year_composition_2019}.  The goal of this analysis is to lower the energy threshold of IceTop to reduce the gap with direct measurements.  A new trigger was introduced in May 2016 to collect small events in the more densely instrumented central area of the array. In this way the threshold of IceTop has been reduced to \unit[250]{TeV}. Events are reconstructed using a random forest regression~\cite{Breiman:2001hzm, James:2014} process trained on simulation.
    
    This paper is divided into five sections. Section~\ref{section_detector} describes the IceTop detector and the new trigger implemented to collect low energy cosmic ray air showers. Next, we describe the experimental and simulated data in Sec.~\ref{section_data}. Section~\ref{section_analysis} describes the reconstruction of air showers based on machine learning and reports the result of the all-particle cosmic ray energy spectrum. This section also describes details of the analysis, including quality cuts, the unfolding method, pressure correction, and systematic uncertainties. Section~\ref{section_result} summarizes the results. An appendix includes tables of systematic uncertainties and numerical values of the spectrum, as well as plots that illustrate the ability of the Monte Carlo to reproduce details of the detector response.

\section{Detector}\label{section_detector}
\begin{figure}
	\centering 
	\includegraphics[height=8.2cm]{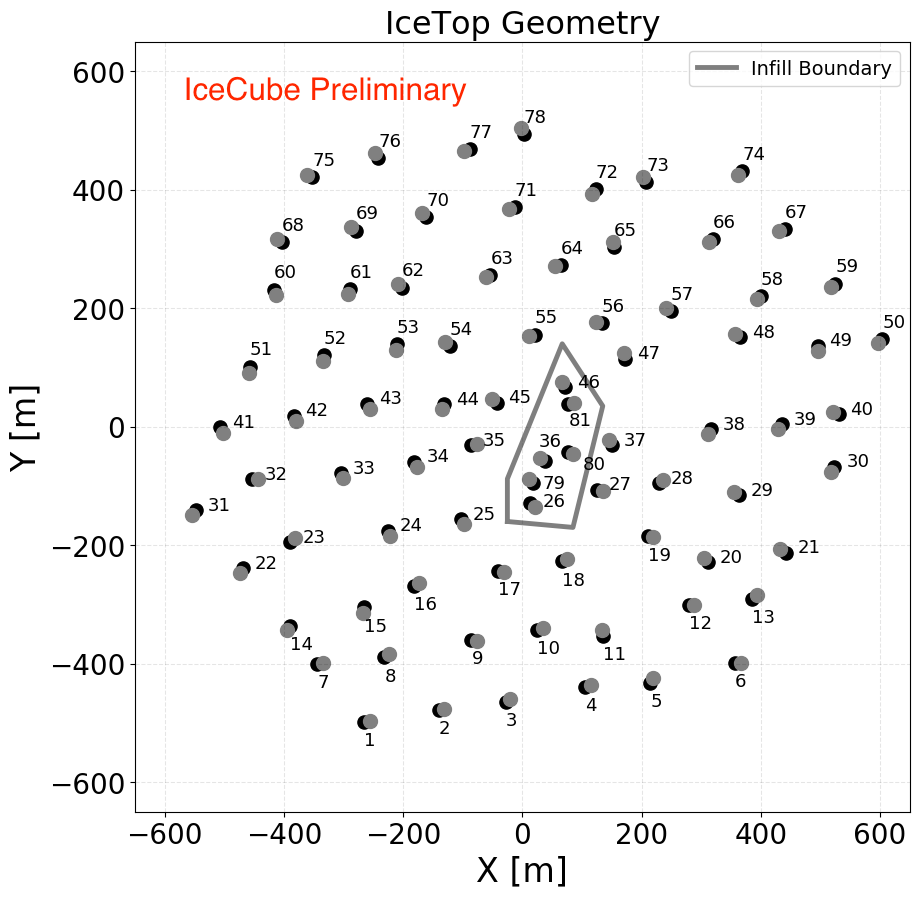}
	\caption{IceTop geometry with positions of all tanks. The marked boundary in the center includes the six stations used to define the two-station trigger.}
	\label{it_geometry}
    \end{figure}

    IceTop is the surface component of the IceCube Neutrino Observatory~\cite{icecube_detector, icetop_detector} at the South Pole. The IceTop array, at an altitude of \unit[2835]{m} above sea level, consists of 162 tanks filled with clear ice distributed in 81 stations spread over an area of \unit[1]{km$^2$} as shown in Fig.~\ref{it_geometry}. Each station has two tanks separated by 10 m. Having two tanks in a station allows for the selection of a subset of events in which both tanks have signal above threshold (hereafter called a ``hit"), suppressing the background of small showers hitting only one tank ($\sim$2 kHz). Details of signal thresholds and other aspects may be found in the IceTop detector paper~\cite{icetop_detector}. Stations are arranged in a triangular grid with typical spacing of 125 m. In addition, IceTop has a dense infill array where the distance between stations is significantly smaller. In this analysis, we refer to stations 26, 36, 46, 79, 80, and 81 as infill stations. 

    Data collected by IceTop are primarily used to measure the cosmic ray energy spectrum~\cite{it26_spectrum,ic40_spectrum,bhaktiyar,IceCube_3year_composition_2019}, to study the mass composition of primary particles~\cite{IceCube_3year_composition_2019}, and to calibrate the IceCube detector~\cite{Bai:2007zzm}. Reference~\cite{IceCube_3year_composition_2019} describes an analysis using 3 years of data (henceforth referred to as “the 3-year IceTop analysis”). IceTop has also been used in searches for PeV gamma rays~\cite{Aartsen:2012gka,pev_gamma_ray}, solar ground level enhancements~\cite{icetop_GLE_2006}, and solar flares~\cite{icetop_GLE_2006}. Another analysis~\cite{JavierMuonLDF} includes single tank hits to identify the component of $\sim$~GeV muons in large showers for studies of primary composition. Finally, IceTop also serves as a partial veto to reduce the background for astrophysical neutrinos~\cite{Tosi:2019nau}. 

     The fundamental detection unit for the IceCube Neutrino Observatory, including IceTop, is the Digital Optical Module (DOM). Each DOM is a glass pressure sphere of \unit[33]{cm} diameter containing hardware to detect light, analyze and digitize waveforms, and communicate with the central data analysis system of IceCube. The photo-multiplier tube (PMT)~\cite{IcecubePMTpaper} is the entry point of light into the data acquisition system (DAQ)~\cite{IceCubeDAQpaper}.  Each IceTop tank contains two DOMs running at different gains.  The DOMs are partially immersed in the clear ice with the PMT facing downward.  Charged particles entering IceTop tanks and photons that convert in the ice produce Cherenkov light that is captured by the PMTs. IceTop DOMs are fully integrated into the IceCube DAQ.  More details of IceTop construction and operation may be found in Ref.~\cite{icetop_detector}.

	\begin{table}[htbp!]
	\centering
	\caption{Four pairs of infill stations and distance between each pair in meters. Each IceTop station has an assigned number from 1 to 81 as shown in Fig.~\ref{it_geometry}.\label{pairtable}}
	\renewcommand{\arraystretch}{1.1}
	\begin{tabular}{cc}
		\hline
		Stations & Distance [m]\\
		\hline
		46, 81 & 34.9  \\ 
		36, 80 & 48.9 \\ 
		36, 79 & 40.7 \\ 
		79, 26 & 41.6 \\ 
		\hline
	\end{tabular}
	\label{station_distance}
	\end{table}

    The standard IceTop trigger, with a requirement of five or more stations with signals, is suitable for detecting cosmic rays in the energy range from PeV to EeV. The six more closely spaced stations in the center of the array (see Fig.~\ref{it_geometry}), are sensitive to cosmic rays with lower energy. The two-station trigger implemented to collect lower energy events is an adaptation of a volume trigger that selects events with hits within a cylinder in the deep array of IceCube. The two-station trigger uses 4 pairs of closely spaced infill stations for which the separation between stations is less than \unit[50]{m} (see Tab.~\ref{pairtable}). (Note that the four pairs are formed with six stations.) 
    
    The trigger condition is satisfied when any of the 4 pairs of infill stations is hit within a time window of \unit[200]{ns}. For this to occur, both tanks at each station of the pair must have signals. Once the trigger condition is fulfilled, the readout window starts \unit[10]{$\mu$s} before the first of the four tank signals and continues until \unit[10]{$\mu$s} after the last. This readout window is sufficient to collect all signals in the entire array of IceTop. The two-station event sample thus includes a subset of events with $\geq$ 5 stations. As a consequence, a subset of the two-station sample can be used to compare the two-station spectrum with the result of the standard IceTop analysis in an overlapping energy region above 2 PeV.
    
 	\begin{figure*}[ht!]
	\centering
	\includegraphics[height=4.5cm]{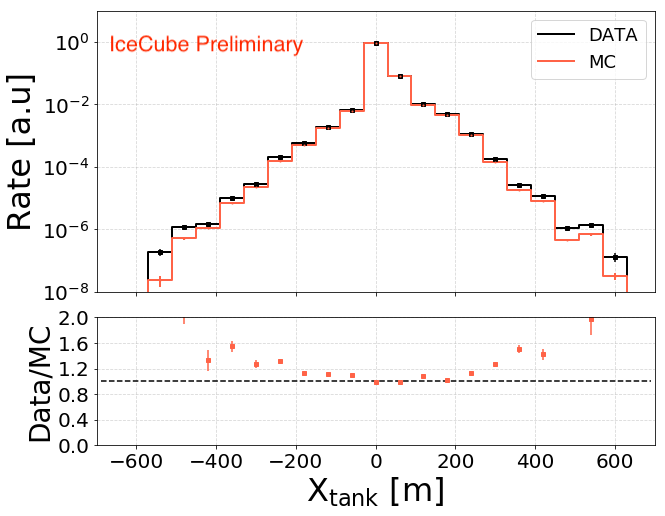}
	\includegraphics[height=4.5cm]{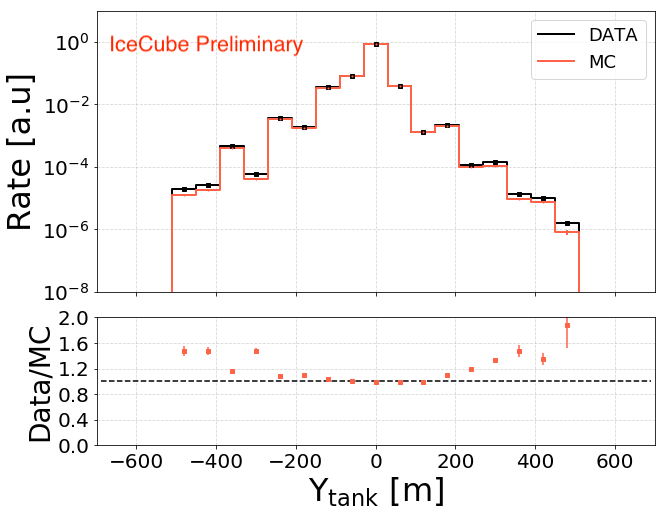}
	\includegraphics[height=4.5cm]{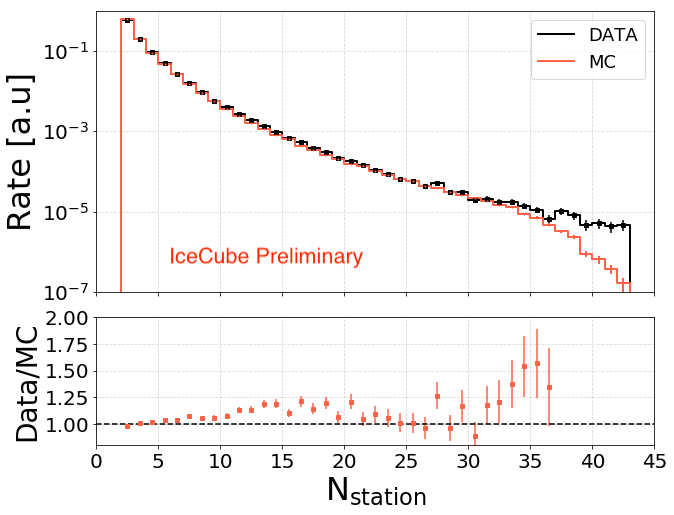}
	\caption{Histograms of x (left) and y (middle) coordinate of tanks hit, and number of stations (right) with signals in data compared with simulations. The differences between experimental data and simulation in the outer regions of coordinates and the higher number of stations arise from the lack of simulation with energy higher than \unit[25.12]{PeV}.}
	\label{fig_tankxy}
	\end{figure*}

\section{Data and Simulations}\label{section_data}
 
    The energy of cosmic rays detected by a ground-based detector is determined indirectly from the signals measured on the ground. In this analysis, energy is reconstructed by a random forest algorithm trained with CORSIKA~\cite{Heck:1998vt} simulations, as described below.
  
\subsection{Data}

    Experimental data were collected from May 2016 to April 2017 (IceCube year 2016) with a livetime of \unit[330.5]{days}. This data set is sufficient to be limited by systematic rather than statistical uncertainty. After all quality cuts, a total of 37,503,350 two-station events survive, of which 7,420,233 lie in the energy region of interest above the \unit[250]{TeV} threshold of reconstructed energy.
    
    The signal in each tank is given by the energy deposited, which is calibrated in units of vertical equivalent muons (VEM).  The VEM is defined as a total integrated charge of the waveform from the energy deposited by a single vertical muon passing through an IceTop tank. Details of signal and timing calibration for the IceTop detector are given in~\cite{icetop_detector}. 
    %------------------------
     \begin{figure}[ht!]
	\centering
	\includegraphics[height=6.5cm]{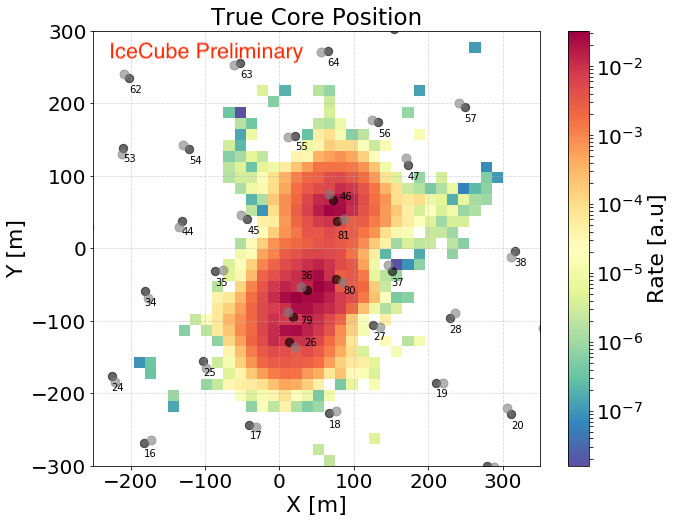}
	\caption{Histogram of true core position of showers after all quality cuts. The position of IceTop tanks is also shown.}
	\label{core_fig}
	\end{figure}

\subsection{Simulations}
    CORSIKA simulations require a representation of the atmospheric density profile and an event generator for the hadronic interactions that make the shower. The atmospheric profile is described below in Sec.~\ref{subsec:pressure_correction}. The hadronic interaction model for the main analysis is Sibyll2.1~\cite{PhysRevD.80.094003}. We also compare results obtained with QGSJetII-04~\cite{qgsjet}.

    CORSIKA simulations of proton, helium, oxygen, and iron primaries ranging from \unit[10]{TeV} to \unit[25.12]{PeV} in energy are used for this analysis. To increase statistics, the same CORSIKA shower is re-sampled multiple times by changing its core position. After re-sampling, there are approximately 100,000 showers for each 0.1 log$_{10}(E/\rm GeV)$ energy bin with zenith angle ($\theta$) up to 65$^\circ$ (except for Helium and Oxygen between \unit[10]{TeV} and \unit[100]{TeV} for which the maximum zenith angle is 40$^\circ$). Events are generated uniformly in sin$^2\theta$ bins. In the zenith region of interest (cos$\theta\geq0.9$, about $\theta<$ 26$^\circ$), there are approximately 24,000 events in each bin of 0.1~log$_{10}(E/\rm GeV)$. Sibyll2.1 is used as the base hadronic interaction model for this analysis so that we can compare the final energy spectrum with the energy spectrum from the 3-year IceTop analysis~\cite{IceCube_3year_composition_2019}. CORSIKA showers with QGSJetII-04 as hadronic interaction model are also produced with 10\% of the statistics compared to that of Sibyll2.1 for a comparative study, to be described in Sec.~\ref{section_flux}. 
    
    Simulations play a vital role in training the random forest reconstruction algorithm. The quality of reconstruction depends on the quality of simulation. There must be a good agreement between simulations and experimental data. To check the quality of simulation, each feature of the experimental data used for the random forest regression is compared to simulation. As an example, Fig.~\ref{fig_tankxy} shows comparison between data and Monte Carlo for three features after the quality cuts described below in Sec.~\ref{section_qualitycuts} have been applied to both data and simulation. The left and middle panels show respectively the distribution of $x$ and $y$ coordinates of tanks with hits. The right panel shows the distribution of the number of tanks with hits. Simulations are normalized to the total number of events in each case. Remaining differences at large distance and for $N_{\rm station}>35$ are a 
    consequence of the lack of simulation above \unit[25.12]{PeV} and do not affect the analysis, which extends only to \unit[10]{PeV}. Similar comparisons for all other features are shown in Appendix~\ref{appendix_data_mc_comparison}. Differences between data and simulation are at the few percent level in the energy region of interest and can therefor be used with good confidence to support the random forest regression for reconstruction of showers.

	In this analysis, each simulated event is weighted based on a 4-component version of the H4a~\cite{Gaisser:2011cc,Gaisser:2013bla} cosmic ray primary composition model. Figure~\ref{core_fig} shows the distribution of true core position of simulated events after weighting by the primary spectrum model and applying the quality cuts described in Sec.~\ref{section_qualitycuts} below. Most events lie within the boundary of the infill area marked 
	in Fig.~\ref{it_geometry}. 

	\begin{table*}[htbp!]
	\centering
	\caption{List of all features that go into random forest regressions and their description. Ref.~\cite{icetop_detector} gives a detailed description of the calculation of the shower center of gravity and direction under the assumption of a plane shower front. See Tab.~\ref{features_list} to know which features are used to reconstruct what air shower's parameter.}
	\renewcommand{\arraystretch}{1.15}
	\begin{tabular}{|p{1.9cm}|p{15.3cm}|}
		\hline
		\multicolumn{1}{|>{\centering\arraybackslash}m{23mm}|}{\textbf{Features}}
    & \multicolumn{1}{>{\centering\arraybackslash}m{13cm}|}{\textbf{Description}}\\
		\hline
		\hline
		X$_{\rm COG}$, Y$_{\rm COG}$ &  X and Y coordinate of shower core's center of gravity.\\ 
		\hline
		$\rm \theta_{\rm plane}$ &  Zenith angle assuming a plane shower front.\\
		\hline
		$\rm \phi_{\rm plane}$ &  Azimuth angle assuming a plane shower front.\\
		\hline
		T$_0$ & Time when shower core assuming plane shower front hits the ground.\\
		\hline
		cos$\rm\theta_{\rm plane}$&  Cosine of $\rm \theta_{\rm plane}$.\\
		\hline
		cos$\rm \theta_{\rm reco}$&  Cosine of reconstructed zenith angle.\\
		\hline
		logNsta & log10 of number of stations hit.\\
		\hline
		logQ$_{\rm total}$ & log10 of total charge deposited in all stations that are hit.\\
		\hline
		Q$_{\rm sum2}$ & Sum of first two highest charges deposited in tanks that are hit.\\
		\hline
		ZSC$_{\rm avg}$ & Average distance of hit tanks from a plane shower front. Absolute value of distance is used to calculate the average. Ideally a ZSC$_{\rm avg}$ is 0 for a vertical shower and is maximal for a horizontal shower.\\
		\hline
		X$_{\rm tank}$, Y$_{\rm tank}$ &  List of X and Y coordinate of hit tanks of each event.\\ 
		\hline
		Q$_{\rm tank}$       &  List of charge deposited on tanks that are hit of each event. \\ 
		\hline
		T$_{\rm tank}$       &  List of hit times on tanks of each event with respect to the first hit time.\\
		\hline
		R$_{\rm tank}$       &  List of distance of hit tanks from the reconstructed shower core of each event. Each distance is divided by \unit[60]{m}.\\
		\hline
	\end{tabular}
	\label{features_description}
	\end{table*}

\section{Analysis}\label{section_analysis}

    This section describes the machine learning technique and features that are used to reconstruct the core position, zenith angle and energy of two-station events. Quality cuts, iterative Bayesian unfolding, pressure correction, and systematic uncertainties are discussed and the cosmic-ray flux is presented.

 	\begin{table}[htbp!]
	\centering
	\caption{Features that go into random forest regressions while training and predicting shower's core position (x and y coordinate), zenith angle, and energy. Four separate random forest regressions are used in this analysis.}
	\renewcommand{\arraystretch}{1.15}
	\begin{tabular}{|l|l|}
		\hline
		\multicolumn{1}{|>{\centering\arraybackslash}m{23mm}|}{\textbf{Reconstructed Variable}}
    & \multicolumn{1}{>{\centering\arraybackslash}m{6cm}|}{\textbf{Features Used}}\\
		\hline
		\hline
		x-coordinate & X$_{\rm COG}$, Y$_{\rm COG}$, X$_{\rm tank}$, Y$_{\rm tank}$,      Q$_{\rm tank}$, \\ 
		& cos$\rm\theta_{\rm plane}$, logNsta \\ 
		\hline
		y-coordinate & X$_{\rm COG}$, Y$_{\rm COG}$, X$_{\rm tank}$, Y$_{\rm tank}$, Q$_{\rm tank}$, \\ 
		& cos$\rm\theta_{\rm plane}$, logNsta \\  
		\hline
		Zenith & $\rm \theta_{\rm plane}$, T$_{\rm tank}$, ZSC$_{\rm avg}$, Q$_{\rm tank}$, T$_{\rm 0}$, logNsta, \\
		  &     $\rm \phi_{\rm plane}$, X$_{\rm COG}$, Y$_{\rm COG}$ \\ 
		\hline
		Energy & Q$_{\rm tank}$, R$_{\rm tank}$, cos$\rm\theta_{\rm reco}$, logQ$_{\rm sum2}$, logNsta, \\
		   &     logQ$_{\rm total}$\\
		\hline
	\end{tabular}
	\label{features_list}
	\end{table}

\subsection{Reconstruction}\label{subsec:reconstruction}

	Four separate random forest (RF) regressions are used for shower reconstruction. Two RFs are used to reconstruct the $x$ and $y$ coordinates of the shower core. A third RF is used to reconstruct the zenith angle, and then a fourth RF is used to determine the shower energy. The azimuthal angle is calculated from a fit of arrival times to a plane shower front. All features used in the regressions are defined in Tab.~\ref{features_description}, and Tab.~\ref{features_list} gives the breakdown of which features are used for each reconstructed quantity.

	For reconstruction of the $x$ coordinate of core position, simulated data are randomly shuffled and divided into two halves to avoid using the same simulated data for training and prediction. If the first half is used for training, the model it generates is used to predict the second half and vice-versa. The predictive capability of machine learning depends on the quality of input data. Events with most of the charge in one or two tanks cannot be reconstructed well, and are therefore omitted. Two-station events in which sum of the two highest charge is more than 95\% of the total charge are removed. The $y$ coordinate of core position is reconstructed by repeating the process used for reconstructing $x$ coordinate. 
	
	Procedures for reconstruction of the coordinates of the core and for reconstruction of zenith angle are similar. The same quality cuts and procedure are implemented to train and to predict zenith angle. The only difference is the features used. Random forest regression from Spark~\cite{spark} is used to reconstruct zenith angle as well as the x and y coordinates of core position. 
	
	Once the reconstruction of $x$ and $y$ coordinates of the core position and the zenith angle is completed, reconstruction of energy is performed. Reconstructed quantities from these initial steps are among the input features for the reconstruction of energy. In addition to the cuts used for core position and zenith angle reconstruction, events are required to have their maximum charge on one of the infill stations given in Tab.~\ref{pairtable}. Also, events with cos$\theta_{reco}<0.8$ are removed. During training for energy reconstruction, events are weighted to the relative abundances of the primary nuclei in the H4a composition model. Energy reconstruction is performed using random forest regression from the Scikit-Learn package~\cite{scikit_learn} with the same random forest regression as in Spark. The only difference is the ability of Scikit-Learn to weight an input event during training by a realistic composition model that Spark lacks. The input weight of events based on H4a composition model during training removes an energy-dependent bias on the reconstructed energy.
    
    ``Ground" is defined at a fixed elevation (\unit[2829.93]{m} above sea level, +\unit[1946]{m} in the IceCube coordinate system), which is the average elevation of all DOMs of two-station events. This elevation is used both for data and simulation. $ZSC_{\rm avg}$ is the average perpendicular distance of DOMs with signals from the plane shower front when the shower core reaches the ground. It is higher for inclined showers and approximately zero for vertical showers. It is given by
	\begin{equation}
	ZSC_{\rm avg} = \frac{\sum_{i=1}^{n}|z_i|}{n},
	\label{zsc_eqn}
	\end{equation}
	where $i$ runs over $n$ hit tanks and $z$ is the position of a tank in the shower coordinate system.
    
    We arrange the position of tanks based on their corresponding charges, largest to smallest, for each air shower. These lists are denoted by $X_{\rm tank}$ and $Y_{\rm tank}$ representing $x$ and $y$ coordinates, respectively. The list of charges in descending order is denoted by $Q_{\rm tank}$. $T_{\rm tank}$ denotes the list of times at which tanks have been hit for each event and is arranged in ascending order. The time of the first hit of an event is subtracted from all hits such that the time used is with respect to the first hit. $R_{\rm tank}$ is defined as a list of the distance from the shower core of each tank that has been hit. Reconstructed shower core using random forest regression is used to calculate $R_{\rm tank}$. The distance of each tank within the array arranged in descending order is divided by a reference distance of \unit[60]{m}. These tank lists are arranged on an individual basis in a particular order based on existing knowledge to increase their predicting ability. For example, the shower core is closer to the tank with the highest charge, hence $X_{\rm tank}$ and $Y_{\rm tank}$ are arranged based on the amount of deposited charge. 

    For each event, these tank lists contain information from n tanks with signals. The minimum value that $n$ can have is 4 and it can in principle increase up to 162. For the energy region of interest, however, information from the first 35 hits is enough to reconstruct shower core position, zenith angle, and energy with nearly 100\% feature importance. Random forest regression becomes computationally expensive as the number of features increases. Therefore, the number of items per event in each list is truncated to 35 from 162. If the number of tanks (n) that have been hit is less than 35, then the remaining 35-n slots of the list are filled with 0 for $X_{\rm tank}$, $Y_{\rm tank}$, $Q_{\rm tank}$, and $R_{\rm tank}$. The remaining slots of $T_{\rm tank}$ are filled with the last relative time.
    
\begin{figure}
	\centering 
	\includegraphics[height=5.4cm]{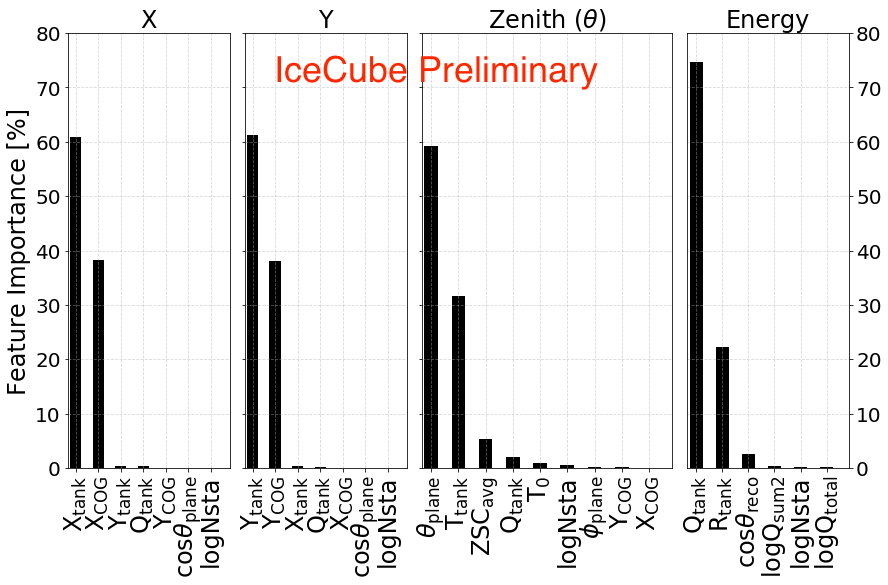}
	\caption{Feature importance of all features used to predict $x$ and $y$ coordinate of core position, zenith angle, and energy. Lists of coordinates of hit tanks ($X_{\rm tank}$, $Y_{\rm tank}$) have the highest feature importance for core position. The zenith angle assuming a plane shower front ($\theta_{\rm plane}$) has the highest feature importance for zenith angle. The list of charge on hit tanks ($Q_{\rm tank}$) has the highest feature importance for energy.}
	\label{fi_plot}
    \end{figure}

    A mean decrease in impurity (MDI) method is used to calculate a feature's importance while predicting core position and zenith angle. MDI and other techniques to calculate feature importance are discussed in~\cite{Louppe_feature_importance, James:2014}. For calculating the importance of features for energy reconstruction, the permutation importance method is implemented. A feature has a list of values, one per event. These values are randomly shuffled so that they no longer belong to their event.  This process is repeated for one feature at a time and feature importance is calculated before and after shuffling. The difference gives importance of that feature. The importance of features listed in Tab.~\ref{features_list} are shown in Fig.~\ref{fi_plot}.
    
	\begin{figure*}[ht!]
	\centering
	\includegraphics[height=5.7cm]{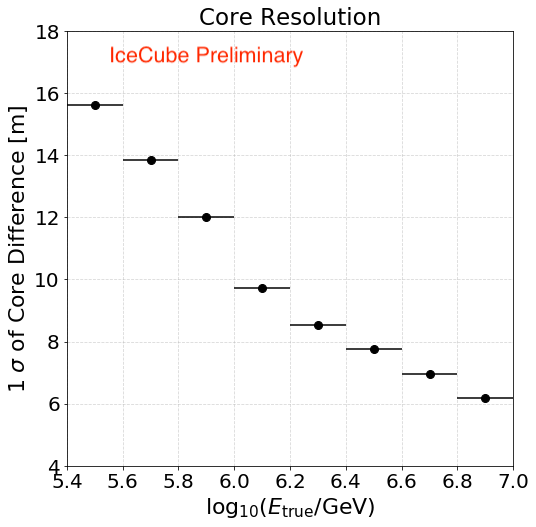}
	\includegraphics[height=5.7cm]{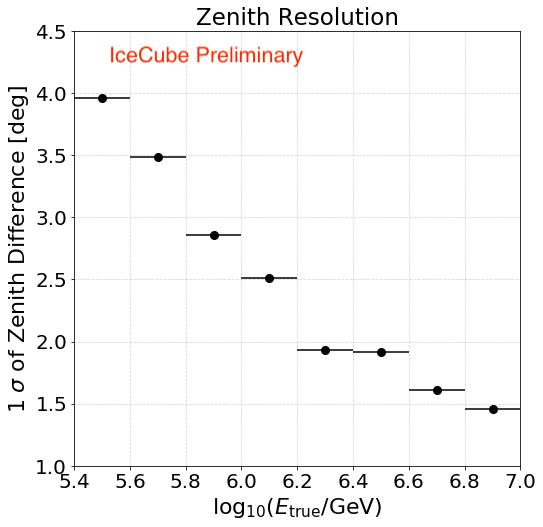}
	\includegraphics[height=5.7cm]{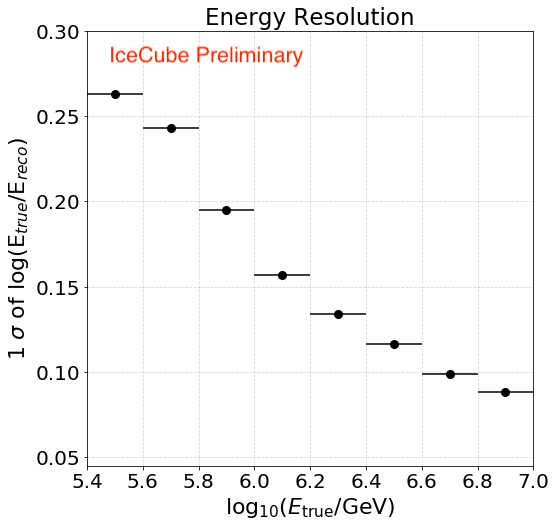}
	\caption{Left: Core resolution in meters; Middle: zenith resolution in degrees; Right: energy resolution in unit-less quantity. See Tab.~\ref{resolution_table} for resolution values.}
	\label{resolution_plot}
	\end{figure*}

 \subsection{Quality cuts}\label{section_qualitycuts}

    Only well-reconstructed events are used to obtain the energy spectrum. Quality cuts are used to remove events with bad reconstruction to reduce error and to improve accuracy. 
    The passing rate of events for a cut is the percentage of events surviving that cut and all previous cuts. Events that pass the two-station trigger conditions have a passing rate of 100\% by definition. The following cuts are applied 
    to the simulated and the experimental data to select events for construction of the energy spectrum:
    \begin{itemize}
	\item Events must have the tank with the highest charge inside the infill boundary. This cut is designed to select events with shower cores inside or near the infill boundary. Passing rate for this cut is 89.5\%.
	\item Events must have cosine of reconstructed zenith angle greater than or equal to 0.9. These events have higher triggering efficiency and are better reconstructed. Passing rate for this cut is 48.1\%.
	\item Events with most of the energy deposited only in few tanks are removed, as they are likely to be poorly reconstructed. This cut requires the largest charge to be less than or equal to 75\% of the total charge and the sum of the two largest charges less than or equal to 90\% of the total charge. Passing rate for this cut is 36.8\%.
    \end{itemize} The simulation used for this low energy analysis extends only to \unit[25.12]{PeV} (log$_{10}$($E$/GeV)=7.4). We have determined that events with true energies above \unit[25.12]{PeV} can be removed by excluding from the data sample events with more than 42 stations hit and excluding events with a total charge greater than \unit[10$^{3.8}$]{VEM}. We found good agreement between data and Monte Carlo after making these two cuts. We also excluded events with a total charge less than \unit[0.63]{VEM} to remove events due to background noise. The number of events removed with these cuts is negligible ($\approx$ 0.003\%). The final cuts listed here are somewhat stronger than those used during reconstruction~(\ref{subsec:reconstruction}) to account for resolution
    near parameter boundaries.  For example, the cut on zenith angle during energy reconstruction is $\cos\theta >0.8$.

    Figure~\ref{resolution_plot} shows core resolution, zenith resolution, and energy resolution. The core resolution is about \unit[16]{m}, the zenith resolution is about $4^\circ$, and the energy resolution is about 0.26 for the lowest energy bin ($5.4 < \log_{10}(E/{\rm GeV})< 5.6$). All three resolutions improve as energy increases. Resolutions for each energy bin are given in Tab.~\ref{resolution_table} of Appendix~\ref{appendix_result}.
    
\subsection{Bayesian Unfolding}
    
    One of the challenges that a ground-based detector faces is to determine the true energy distribution ($C$, Cause) from the reconstructed energy distribution ($E$, Effect). Iterative Bayesian unfolding~\cite{DAgostini1995, dagostini_unfolding} is used to take energy bin migration into account and to derive the true from the reconstructed energy distribution. It is implemented via a software package called pyUnfolding~\cite{pyunfolding}. This package also calculates and propagates error in each iteration.

\begin{figure}[ht!]
	\centering
	\includegraphics[height=7.5cm]{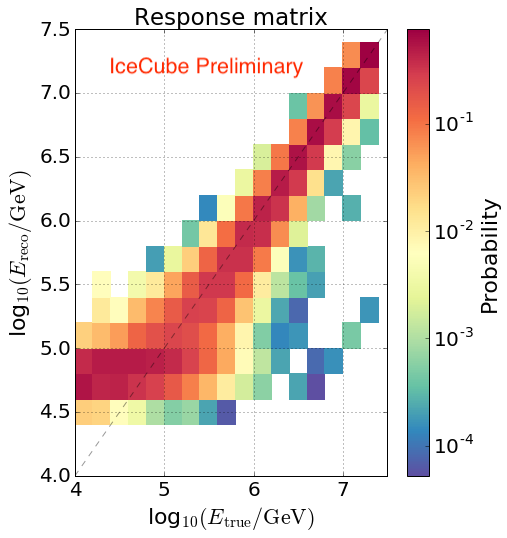}
	\caption{Response Matrix calculated from simulation with Sibyll2.1 as the hadronic interaction model. An element of a response matrix is a fraction of events in true energy bin distributed over the reconstructed energy bin. In Bayes theorem of Eq.~\ref{bayestheorem}, $P(E|C)$ represents a response matrix.}
	\label{fig_response_matrix}
    \end{figure}

    To unfold the energy spectrum, the detector response to an air shower is required. The response is determined from simulations as the probability of measuring a reconstructed energy given the true primary energy. The information stored in a response matrix is shown in Fig.~\ref{fig_response_matrix}. Inverting the response matrix to get a probability of measuring true energy given reconstructed energy would lead to unnatural fluctuations. Therefore, Bayes theorem is used iteratively to get the true distribution from the observed distribution. 

    Bayes theorem is given by 
    \begin{equation}
    P(C_\mu|E_i) = \frac{P(E_i|C_\mu)P(C_\mu)}{\sum_\nu^{n_C}P(E_i|C_\nu)P(C_\nu)},
    \label{bayestheorem}
    \end{equation}
    where $P(C|E)$ is the unfolding matrix, $P(E|C)$ is the response matrix, $n_C$ is the number of cause bins, and $P(C)$ is the prior knowledge of the cause distribution. $P(C)$ is the only quantity that changes in the right-hand side of Eq.~\ref{bayestheorem} during each iteration. The choice of initial prior, $P(C)$, can be any reasonable distribution, like a uniform distribution or a normalized distribution of effect $\phi(E)/\sum_i^{n_E}\phi(E_i)$ where $n_E$ is the number of effect bins. The conventional choice to minimize bias is Jeffreys' Prior~\cite{jeffreys_prior}, given by $$P_{\rm Jeffreys}(C_\mu) = \frac{1}{\log_{10}(C_{\rm max}/C_{\rm min})C_\mu}$$.

    Each iteration produces a new unfolding matrix $P(C|E)$. $P(C_\mu|E_i)$ represents the probability that an effect $E_i$ is a result of cause $C_\mu$. If the distribution of effect $\phi(E)$ is known then the updated knowledge of the cause distribution is given by
    \begin{equation}
    \phi(C_\mu) = \frac{1}{\epsilon_\mu}\sum_i^{n_E}P(C_\mu|E_i)\phi(E_i),
    \label{phicmu}
    \end{equation}
    with $\epsilon_\mu = \sum_j^{n_E}P(E_j|C_\mu)$, where in this analysis $\epsilon_\mu=1$.  $\phi(C_\mu)$ in Eq.~\ref{phicmu} is used to calculate an updated prior. The updated prior is given by
    \begin{align*}
    P(C_\mu) = \frac{\phi(C_\mu)}{\sum_\nu\phi(C_\nu)},
    \end{align*}
    which is then used as a new prior in Eq.~\ref{bayestheorem} for the next iteration. The unfolding proceeds until a desired stopping criterion is satisfied. In this analysis, a Kolmogorov-Smirnov test statistic~\cite{kolmogorov, smirnov1948} of subsequent energy distribution before and after unfolding less than $10^{-3}$ is used as the stopping criterion. It is reached in the twelfth iteration.

    \begin{figure}[ht!]
	\centering
	\includegraphics[height=7cm]{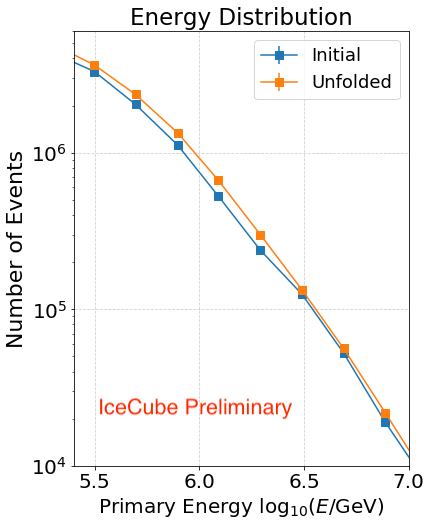}
	\caption{Energy distribution before and after iterative Bayesian unfolding. Blue is the reconstructed energy distribution and orange is the final unfolded energy distribution.}
	\label{unfolded_ene_dist}
    \end{figure}

    Using a new cause distribution ($\phi(C)$) to calculate the next prior can propagate error, if any, in each iteration which might cause erratic fluctuations on the final distribution. To regularize the process and to avoid passing an unphysical prior, the logarithm of the cause distribution $(\phi(C))$ is fitted with a third degree polynomial in each iteration except for the final distribution. The final unfolded energy distribution is used to calculate the cosmic ray flux. The initial reconstructed energy distribution ($\phi(E)$ in Eq.~\ref{phicmu}) and the final unfolded energy distribution are shown in Fig.~\ref{unfolded_ene_dist}.

\subsection{Pressure Correction}\label{subsec:pressure_correction}
    
    The rate of two-station events fluctuates with changes in atmospheric pressure. If pressure increases, the rate decreases because the shower is attenuated by having to go through more mass above the detectors and vice-versa.  At least two factors contribute to the rate fluctuations. At higher pressure, the signal at ground for a given energy is smaller. In addition, the shower is more spread out, decreasing the trigger probability. If the average pressure during which data were taken is not equal to the pressure of the atmospheric profile used in the simulation, then the final flux must be corrected to account for the difference in the atmospheric pressure between data and simulation. This correction is applied to the full data sample after Bayesian unfolding.

    \begin{figure}[ht!]
	\centering
	\includegraphics[height=6.5cm]{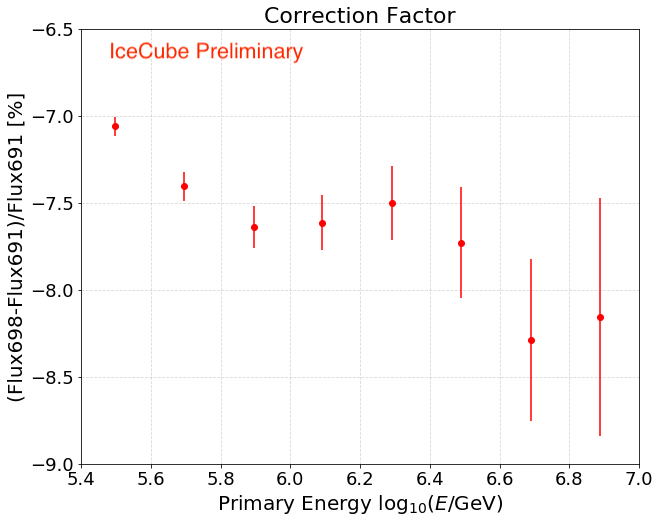}
	\caption{Percentage deviation of cosmic rays flux when atmospheric pressure is $\sim$\unit[698]{g/cm$^2$} from the flux when pressure is $\sim$\unit[691]{g/cm$^2$}. This deviation is used as the  correction factor to correct the final flux. The error on the correction factor is used as the systematic uncertainty due to pressure difference between average pressure of 2016 South Pole atmosphere and pressure due to atmosphere profile used in simulation.}
	\label{sys_pressure}
    \end{figure}

    %AMRC data: http://www.bartol.udel.edu/~takao/icetop/climate/surface/index_2016.html
    The average pressure at the South Pole during data-taking was \unit[678.27]{hPa} (data obtained from Antarctic Meteorological Research Center). This converts to \unit[691.16]{g/cm$^2$} using a conversion factor 1.019 (g/cm$^2$/hPa). 
    
   The atmospheric density variation is modeled in CORSIKA with 5 layers. In each layer except the highest, the overburden $T(h)$ of the atmosphere is given by  
    \begin{equation}
        T(h) = a + b\exp\left(-\frac{h}{c}\right),
    \end{equation}
   where $h$ is the altitude from sea level.
   The average April atmosphere was used for this analysis.  The lowest layer extends up to \unit[7.6]{km}, for which the parameters are $a=-69.7259$ g/cm$^2$, $b=1111.7$ g/cm$^2$, and $c=766099$ cm.  With these parameters for IceTop at an altitude of \unit[2835]{m} above sea level, $T(h)$ is  \unit[$698.12$]{g/cm$^2$}. This is $\sim$1\% larger than the average pressure for the period of data-taking ($698.12/691.1$). 

   To address this problem, we selected a shorter data period (08 January 2017 to 28 April 2017) during which
   the average pressure was \unit[698.12]{g/cm$^2$}, the same as that used in the simulation. The flux from this subset of data is calculated and compared with the flux for the entire data-taking period. The flux decreases with an increase in pressure and this decrease must be corrected. The correction factor is shown in Fig.~\ref{sys_pressure} and tabulated in Tab.~\ref{pressure_correction_table}, Appendix~\ref{appendix_result}. The correction shifts the flux down. Errors on the correction factors due to pressure difference are included in the estimation of the flux systematic uncertainties.%

\subsection{Systematic Uncertainties}
\begin{figure}[ht!]
	\centering
	\includegraphics[height=6cm]{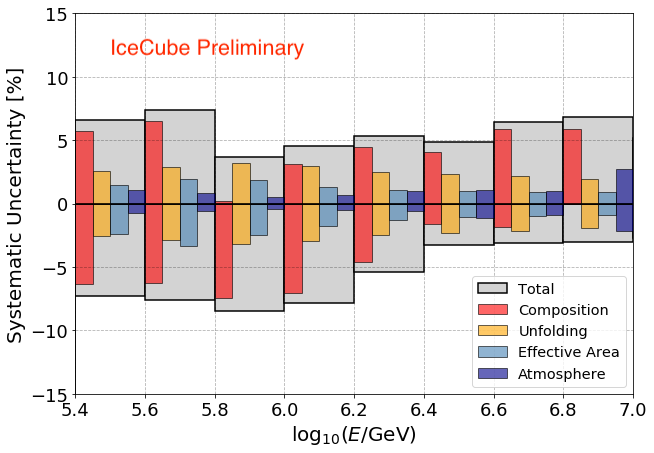}
	\caption{The plot shows the individual systematic uncertainties for each energy bin. The total systematic uncertainty is the sum of individual uncertainties added in quadrature.}
	\label{total_sys_uncertainty}
    \end{figure}

    The major systematic uncertainties, excluding those due to hadronic interaction models, are those due to the composition, the unfolding method, the effective area, and the atmosphere.  Both individual and total systematic uncertainties are shown in Fig.~\ref{total_sys_uncertainty}. The systematic uncertainties due to the hadronic interaction models are not considered, but results assuming Sibyll2.1 and QGSJetII-04 as hadronic interaction models are presented separately.
    
    To estimate the uncertainty due to composition, the H4a model~\cite{Gaisser:2011cc,Gaisser:2013bla} (with four groups of nuclei) is used as the base composition model. The three population fit (Table III of GST~\cite{Gaisser:2013bla}), GSF~\cite{Dembinski:2017zsh}, and a version of the Polygonato model~\cite{Hoerandel:2004gv} are used as alternate models.  (The original Polygonato model is extended by the addition of the second Galactic population B from H4a at high energy and modified at low energy to combine nuclei into groups as in H4a.) Since all these models are viable options for composition, the flux for each model is calculated using the same response matrix, and the percentage deviation of the flux from the model for each energy bin is measured. Additionally, the fractional difference between fluxes calculated for two extreme zenith bins (0.8$\geq$~cos$\theta\geq$0.9 and 0.9$\geq$~cos$\theta\geq$1) is used to calculate the percentage deviation of flux due to composition systematics as done in~\cite{bhaktiyar}. The maximum spread of all deviations is used as the uncertainty due to composition.

    The pyUnfolding software package calculates the systematic uncertainty due to unfolding at the end of each iteration. The uncertainty arises from the limited statistics of the simulated data set. Evolution of systematic uncertainty after each iteration is saved. In this study, we need 12 iterations before reaching the termination criterion. The systematic uncertainty for the twelfth iteration is used as the systematic uncertainty from the unfolding procedure.
 
    \begin{figure}[ht!]
	\centering
	\includegraphics[height=6cm]{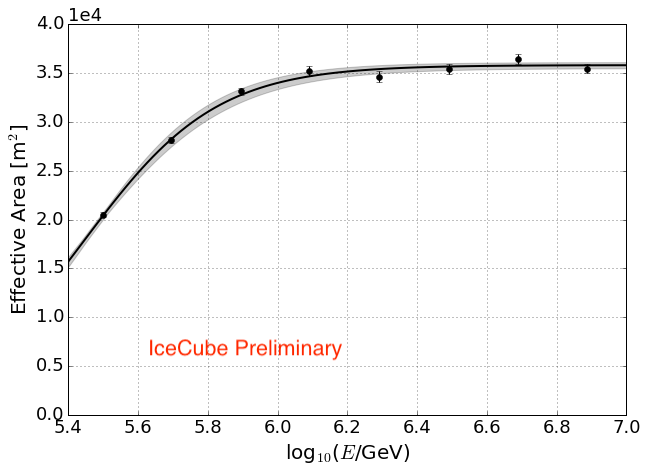}
	\caption{Effective area calculated using MC generated with H4a composition model and Sybill 2.1 hadronic interaction model. A sigmoid function is used to fit the effective area.}
	\label{effective_area_plot}
    \end{figure}
   
    The cosmic-ray flux is the ratio of the number of reconstructed events per logarithmic bin of energy divided by the product of the effective area, exposure time and solid angle.  Uncertainties in exposure time and solid angle are small compared to the uncertainties from primary composition and unfolding.  Effective area is determined from simulation as the sampling area used in the simulation multiplied by the efficiency as a function of primary energy.  Efficiency is the ratio of the number of events that survive all quality cuts to the number of simulated events.  The points in Fig.~\ref{effective_area_plot} give the effective area for each bin of log(E). The effective area is fitted to an energy-dependent function of the form:
    \begin{equation}
        A_{\rm eff}(E) = \frac{p_0}{1+{\rm e}^{-p_1 (\log E - p_2)}}.
        \label{sigmoid_like_function}
    \end{equation}
    where $p_0$, $p_1$, and $p_2$ are free parameters. The parameters of the fit contain uncertainties that are used to estimate the systematic uncertainty in the effective area. A band around the effective area fit is shown in Fig.~\ref{effective_area_plot} after accounting for all errors on the parameters. Taking the upper and lower boundary of the band, the flux is calculated and the difference in the flux is used as the systematic uncertainty due to effective area.

    The correction factor to account for the atmospheric pressure difference between data and simulation is shown in Fig.~\ref{sys_pressure}. The uncertainty on the correction factor is used as the systematic uncertainty due to pressure correction. Also, the difference in flux due to different temperatures for constant pressure is used as the systematic uncertainty due to temperature and is less than 2\%. These two uncertainties are added and the summation is used as the systematic uncertainty due to the atmosphere.

    Snow accumulates over the tanks and the effect of its absorption is accounted for in the simulation, as described in~\cite{IceCube_3year_composition_2019}. Different snow heights for data and simulation would cause a systematic shift in the low energy spectrum. Experimental data used in this analysis is from May 2016 to April 2017 and the snow height used for simulations is from October 2016 in the middle of the data sample. Annual snow accumulation at the South Pole averages \unit[20]{cm}, so on average differs by less than $\pm$\unit[10]{cm} ($\sim$\unit[4]{cm} water equivalent) over the period of data taking.  This range is symmetric about the snow depth used in the simulation and is less than half a percent of the total atmospheric overburden, so no systematic error is assigned.  
    
    VEM calibration occurs monthly. Systematic uncertainty arising from VEM calibration has been studied and is only $\sim$0.3\%. Therefore, systematic uncertainty due to VEM calibration is ignored. 

    The statistical uncertainty of the energy spectrum is small due to the large volume of data. The systematic uncertainty from the composition assumption is the largest, while the systematic uncertainties from the unfolding method, effective area, and atmosphere are relatively small. The `total systematic uncertainty' is calculated by adding individual contributions in quadrature and is larger than the statistical uncertainty. The total systematic uncertainty for each energy bin is tabulated in Tab.~\ref{sys_total_table} of Appendix~\ref{appendix_result}.
 
\subsection{Flux}\label{section_flux}
    Once the core position, direction, and energy of air showers are reconstructed, and the effective area is known, the flux is calculated. The binned flux is given by 
    \begin{equation}
    J(E) = \frac{\Delta N(E)}{\Delta {\rm ln}E \pi(\cos^2\theta_1 - \cos^2\theta_2) A_{\rm eff} T},
    \label{flux_equaiton_final}
    \end{equation}
    where $\Delta N(E)$ is the unfolded number of events with energy per logarithmic bin of energy in time $T$, [$\theta_1, \theta_2$] is the observed zenith range, and $A_{\rm eff}$ is the effective area. The effective area for IceTop events with $\cos\theta\geq 0.9$ is shown in Fig.~\ref{effective_area_plot} and is used to calculate the flux. The livetime ($T$) is \unit[28548810]{s} (\unit[330.5]{days}), $\Delta \log_{10}E$ is 0.2, and $\cos\theta_1=1.0$ and $\cos\theta_2=0.9$.

    The all-particle cosmic ray flux is calculated using Eq.~\ref{flux_equaiton_final} in the energy range \unit[250]{TeV} to \unit[10]{PeV}. The calculated flux is corrected for pressure difference using the correction factors shown in Tab.~\ref{pressure_correction_table} of Appendix~\ref{appendix_result}. The final flux is then compared with the higher energy measurement of IceCube~\cite{IceCube_3year_composition_2019} in the left plot of Fig.~\ref{cr_flux}. Table~\ref{cr_flux_table} in Appendix~\ref{appendix_result} tabulates the result of the IceTop low energy spectrum analysis.

    The effect of the hadronic interaction model is not included in the total systematic uncertainty. Instead, the same analysis steps were repeated using simulation with QGSJetII-04 as the hadronic interaction model. The statistics of the simulation for the analysis with QGSJetII-04 is only 10\% of that for Sibyll2.1 but is sufficient for the comparison between the two models. The right plot of Fig.~\ref{cr_flux} shows the comparison between fluxes assuming Sibyll2.1 and QGSJetII-04 as hadronic interaction models and their ratio. The flux assuming QGSJetII-04 is comparable with the flux assuming Sibyll2.1 at the lower energy region and is around 20$\%$ lower above the knee. Above a PeV, the proton cross section in Sibyll2.1 increases with energy somewhat faster than that in QGSJetII-04~\cite{Ostapchenko:2010vb}, and this may contribute to the lower flux above the knee. Results assuming QGSJetII-04 as the hadronic interaction model are tabulated in Tab.~\ref{cr_flux_table_qgsjet} of Appendix~\ref{appendix_result}.

    \begin{figure*}[ht!]
	\centering
	\includegraphics[height=7cm]{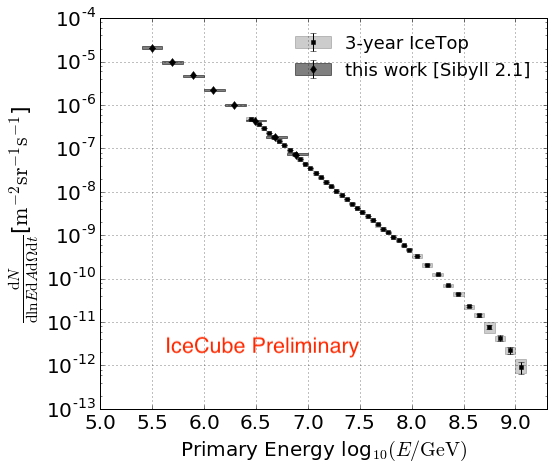}
	\includegraphics[height=7cm]{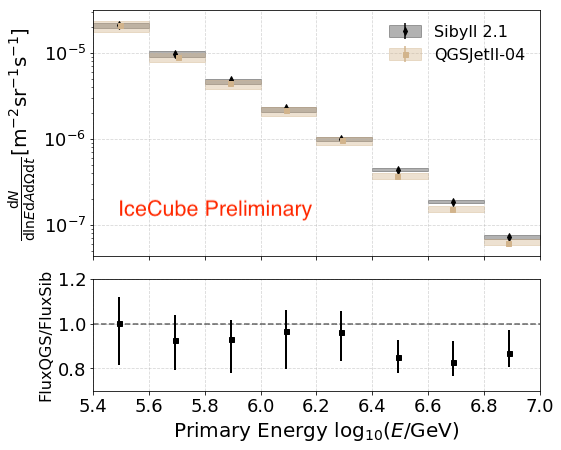}
	\caption{Left: The all-particle cosmic ray energy spectrum using IceTop 2016 data compared to the IceTop measurement at high energy~\cite{IceCube_3year_composition_2019}. Right: The all-particle cosmic ray energy spectra using simulations with Sibyll2.1 and QGSJetII-04 as hadronic interaction models. The same analysis as with Sibyll2.1 was repeated with QGSJetII-04. The shaded region in both plots indicates the systematic uncertainties.}
	\label{cr_flux}
    \end{figure*}

    \begin{figure*}[ht!]
	\centering
	\includegraphics[height=8.5cm]{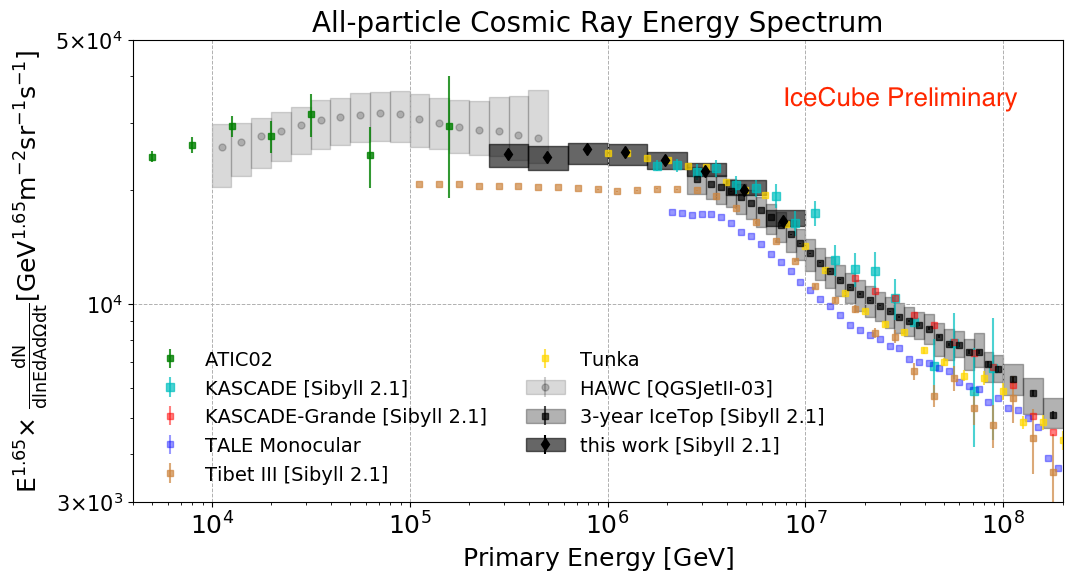}
	\caption{Cosmic ray flux using IceTop 2016 data scaled by $E^{1.65}$ and compared with flux from other experiments. This analysis and HAWC's energy spectrum analysis~\cite{HAWC2017} use different hadronic interaction models. The shaded region indicates the systematic uncertainties. The energy spectra from ATIC-02~\cite{atic_spectrum}, KASCADE~\cite{kascade_spectrum}, KASCADE-Grande~\cite{Bertaina:2015fnz}, TALE~\cite{tale_spectrum}, Tibet III~\cite{Amenomori} and Tunka~\cite{PROSIN201494} are also plotted to compare with the energy spectrum from this analysis.}
	\label{cr_flux_many_experiments}
    \end{figure*}
    
\section{Result and Discussion}\label{section_result}
    The principal result of this paper is the all-particle cosmic-ray spectrum from \unit[250]{TeV} to \unit[10]{PeV}, covering the full knee region with a single measurement. 
    Figure~\ref{cr_flux_many_experiments} makes clear the behavior of the spectrum through the knee region, with an integral slope of $1.65$ below a PeV and a gradual steepening between \unit[2]{PeV} and \unit[10]{PeV}. Lowering the energy threshold from $\sim$\unit[2]{PeV} to \unit[250]{TeV} reduces the gap between IceTop and direct measurements. This is the first result using the new two-station trigger. Several measurements from other experiments with their statistical uncertainties are compared with the result of this analysis in Fig.~\ref{cr_flux_many_experiments}.

	The final energy spectrum is somewhat higher than the 3-year IceTop spectrum in the overlap region. These two fluxes are fitted with splines to calculate their percentage differences at each energy bin of the 3-year IceTop analysis up to \unit[10]{PeV}. The flux found from this analysis is within 7.1\% of the 3-year IceTop spectrum. The total systematic uncertainty for the 3-year spectrum is 9.6\% at \unit[3]{PeV} and 10.8\% at \unit[30]{PeV}~\cite{IceCube_3year_composition_2019}. Even though the flux is higher, it is within the systematic uncertainty of the 3-year IceTop energy spectrum analysis. Both analyses use data collected by IceTop, so they share systematic uncertainties related to the detector. However, there are differences in this analysis, such as the treatment of the pressure correction and the unfolding that contribute to the systematics.  Other important differences are in data taking (trigger) and in the use of machine learning for reconstruction.

    Many ground-based detectors have measured the cosmic ray flux in overlapping regions of energy. The range of fluxes, as shown in Fig.~\ref{cr_flux_many_experiments}, reflects systematic uncertainties in the measurements. Since the cosmic ray flux follows a steep power law, a slight difference in energy scale can cause a large difference in the flux. The IceTop low energy spectrum overlaps with the results from HAWC~\cite{HAWC2017} in the lower energy region and with KASCADE~\cite{kascade_spectrum} and Tunka~\cite{PROSIN201494} measurements at higher energies. It is higher than the result from Tibet III~\cite{Amenomori} and TALE~\cite{tale_spectrum}. The low energy spectrum is also compared with a direct measurement from ATIC-02~\cite{atic_spectrum}. Perhaps the most relevant for comparison with the present analysis is Tibet-III, which is a ground-based air shower array at high altitude with closely spaced detectors. They have analyzed their data with Sibyll2.1 as well as with an earlier (pre-LHC) version of QGSJet-01c~\cite{Kalmykov:1993qe}. They compare two composition models, proton-dominated (PD) and heavy-dominated (HD) with the QGSJet interaction model. For Sibyll2.1 they use HD. The Tibet result plotted in Fig.~\ref{cr_flux_many_experiments} is Sibyll2.1 with HD. Based on the Tibet comparison between HD and PD with QGSJet, Sibyll2.1 with PD would be 10\% to 20\% lower, enlarging the difference shown in Fig.~\ref{cr_flux_many_experiments}. The PD composition of Ref.~\cite{Amenomori} is similar to that of H4a used in this paper. It is important to remember that apparent differences between measurements are amplified by the steepness of the spectrum. For example, in the energy region below the knee where the integral spectral index is 1.65 and the ratio of the two measurements shown in Fig.~\ref{cr_flux_many_experiments} is $\approx 1.24$, a difference in energy scale of $12$\% is sufficient to account for the difference. (Explicit formulas to account for energy scale differences are given in Sec.~2.5.2 of \cite{tomsbook}.)
    
    The energy spectrum measured in this analysis fills the gap between the 3-year IceTop spectrum and the HAWC measurements. HAWC, with large, contiguous water Cherenkov tanks, is able to extend its measurement to much lower energy than IceTop and overlaps in energy with direct measurements. Its uncertainty band is larger than that shown for IceTop in part because the uncertainties from hadronic interactions are included for HAWC but not for IceTop. Looking ahead, it is worth noting the effect of updating the cross section of Sibyll2.1 to post-LHC values, which are smaller above a PeV than the cross section in Sibyll2.1. With the smaller cross section, simulated showers will penetrate deeper in the atmosphere, so a given size parameter will correspond to a lower energy, shifting the spectrum down. (For a comparison of $\sigma_{\rm{p-air}}$ between Sibyll2.1 and its post-LHC version, see Ref.~\cite{Engel:2019dsg}.)

    The TALE experiment, using atmospheric fluorescence and Cherenkov radiation, covers an energy range from just below the knee to an energy that overlaps with ultra-high energy cosmic rays in the EeV range.  Because of the steeper spectrum above the knee, the effect of any uncertainty in energy scale is amplified more.  For example, in the energy range $3$--$10$~PeV the integral spectral index of TALE is $2.12$, and a $37$\% difference between the fluxes corresponds to a $16$\% shift in energy scale.

\begin{acknowledgments}
    The IceCube collaboration acknowledges the significant contribution to this paper from the Bartol Research Institute at the University of Delaware. 
	Support of the University of Wisconsin-Madison Automatic Weather Station Program is gratefully acknowledged for the pressure data used in this analysis (NSF grant 1924730).

    The IceCube collaboration also acknowledges support from the following agencies.
    USA {\textendash} U.S. National Science Foundation-Office of Polar Programs,
    U.S. National Science Foundation-Physics Division,
    Wisconsin Alumni Research Foundation,
    Center for High Throughput Computing (CHTC) at the University of Wisconsin-Madison,
    Open Science Grid (OSG),
    Extreme Science and Engineering Discovery Environment (XSEDE),
    U.S. Department of Energy-National Energy Research Scientific Computing Center,
    Particle astrophysics research computing center at the University of Maryland,
    Institute for Cyber-Enabled Research at Michigan State University,
    and Astroparticle physics computational facility at Marquette University;
    Belgium {\textendash} Funds for Scientific Research (FRS-FNRS and FWO),
    FWO Odysseus and Big Science programmes,
    and Belgian Federal Science Policy Office (Belspo);
    Germany {\textendash} Bundesministerium f{\"u}r Bildung und Forschung (BMBF),
    Deutsche Forschungsgemeinschaft (DFG),
    Helmholtz Alliance for Astroparticle Physics (HAP),
    Initiative and Networking Fund of the Helmholtz Association,
    Deutsches Elektronen Synchrotron (DESY),
    and High Performance Computing cluster of the RWTH Aachen;
    Sweden {\textendash} Swedish Research Council,
    Swedish Polar Research Secretariat,
    Swedish National Infrastructure for Computing (SNIC),
    and Knut and Alice Wallenberg Foundation;
    Australia {\textendash} Australian Research Council;
    Canada {\textendash} Natural Sciences and Engineering Research Council of Canada,
    Calcul Qu{\'e}bec, Compute Ontario, Canada Foundation for Innovation, WestGrid, and Compute Canada;
    Denmark {\textendash} Villum Fonden, Danish National Research Foundation (DNRF), Carlsberg Foundation;
    New Zealand {\textendash} Marsden Fund;
    Japan {\textendash} Japan Society for Promotion of Science (JSPS)
    and Institute for Global Prominent Research (IGPR) of Chiba University;
    Korea {\textendash} National Research Foundation of Korea (NRF);
    Switzerland {\textendash} Swiss National Science Foundation (SNSF);
    United Kingdom {\textendash} Department of Physics, University of Oxford.
\end{acknowledgments}

\vfill\eject
\bibliography{paper_main}

\clearpage

\onecolumngrid
\appendix
\section{Resolutions, Correction Factor, Systematic Uncertainty, and the Final Results}

\setlength{\tabcolsep}{8pt}

\label{appendix_result}
    \begin{table*}[htbp!]
	\centering
	\caption{Quality of reconstruction. The first row shows the core resolution in meters. The second row shows the zenith resolution in degrees. The third row shows the energy resolution. This is the tabulation of the numbers in Fig.~\ref{resolution_plot}.}
	\label{resolution_table}
	\renewcommand{\arraystretch}{1.1}
	\begin{tabular}{lcccccccc}
		\hline
		log$_{10}$($E$/GeV) & 5.4-5.6 & 5.6-5.8 & 5.8-6.0 & 6.0-6.2 & 6.2-6.4 & 6.4-6.6 & 6.6-6.8 & 6.8-7.0\\
		\hline
		Core [m] & 15.62 & 13.85 & 12.03 &  9.76 &  8.45 & 7.76 &  6.95 & 6.22 \\
		Zenith [deg] & 3.95 &  3.47 & 2.87 & 2.51 & 1.94 & 1.95 & 1.62 & 1.46 \\
		Energy & 0.26 & 0.24 & 0.20 &  0.16 &  0.14 &  0.12 & 0.10 &  0.09 \\
		\hline
	\end{tabular}
    \end{table*}

    \begin{table*}[htbp!]
	\centering
	\caption{Correction factor on the final flux due to difference in atmospheric pressure between simulation and 2016 data. \label{pressure_correction_table}}
	\renewcommand{\arraystretch}{1.1}
	\begin{tabular}{ccccccccc}
		\hline
		log$_{10}$($E$/GeV) & 5.4-5.6 & 5.6-5.8 & 5.8-6.0 & 6.0-6.2 & 6.2-6.4 & 6.4-6.6 & 6.6-6.8 & 6.8-7.0\\
		\hline
		[\%] & -7.06 &  -7.41 &  -7.64 &  -7.61 &   -7.50 &   -7.73 & -8.29 &  -8.16\\
		\hline
	\end{tabular}
    \end{table*}

    \begin{table*}[htbp!] 
	\centering
	\caption{Total systematic uncertainty after adding individual systematic uncertainty in quadrature. \label{sys_total_table}}
	\renewcommand{\arraystretch}{1.1}
	\begin{tabular}{ccccccccc}
		\hline
		log$_{10}$($E$/GeV) & 5.4-5.6 & 5.6-5.8 & 5.8-6.0 & 6.0-6.2 & 6.2-6.4 & 6.4-6.6 & 6.6-6.8 & 6.8-7.0\\
		\hline
		Low [\%] & 7.27 &  7.64 &  8.45  &  7.85 &  5.43 &  3.26 & 3.12 & 3.07 \\
		High [\%]& 6.54 &  7.39 &  3.70  &  4.53 &  5.35 &  4.88 & 6.43 &  6.84\\
		\hline
	\end{tabular}
    \end{table*}
    
	\begin{table*}[htbp!]
	\caption{Information related to all-particle cosmic ray energy spectrum using two stations events. Sibyll2.1 is the hadronic interaction model assumption. The first column is the energy bin in $\log_{10}(E/{\rm GeV})$. The second column is the number of events in reconstructed energy bins before unfolding. The total number of events in these energy bins is 7,420,233. The third column is the rate of events before unfolding calculated by dividing the second column with livetime. The fourth column is the unfolded rate. The fifth column is the all-particle cosmic ray flux calculated from the unfolded rate.  The remaining columns are the statistical uncertainty, the lower systematic uncertainty, and the upper systematic uncertainty in the flux respectively.}
	\label{cr_flux_table}
	\resizebox{\textwidth}{!}{%
		\renewcommand{\arraystretch}{1}
		\begin{tabular}{lrcccccc}
			\hline
			$\log_{10}(E/{\rm GeV})$ & $N_{\rm events}$& Rate & Unfolded Rate & Flux           & Stat. Err  & Sys Low & Sys High \\
			& & [Hz] & [Hz] & \multicolumn{4}{c}{[$\rm m^{-2}s^{-1}sr^{-1}$]} \\
			\hline
			5.4 - 5.6 & 3,301,846& $1.16\times 10^{-1}$ & $1.27\times 10^{-1}$ & $2.11\times 10^{-5}$ & $1.76\times 10^{-8}$  & $1.53\times 10^{-6}$ & $1.38\times 10^{-6}$ \\
			5.6 - 5.8 & 2,034,816& $7.13\times 10^{-2}$ & $8.25\times 10^{-2}$ & $9.79\times 10^{-6}$ & $1.26\times 10^{-8}$  & $7.48\times 10^{-7}$ & $7.24\times 10^{-7}$ \\
			5.8 - 6.0 & 1,120,920& $3.93\times 10^{-2}$ & $4.70\times 10^{-2}$ & $4.81\times 10^{-6}$ & $9.26\times 10^{-9}$  & $4.06\times 10^{-7}$ & $1.78\times 10^{-7}$ \\
			6.0 - 6.2 & 527,453  & $1.85\times 10^{-2}$ & $2.33\times 10^{-2}$ & $2.26\times 10^{-6}$ & $5.51\times 10^{-9}$  & $1.77\times 10^{-7}$ & $1.02\times 10^{-7}$ \\
			6.2 - 6.4 & 238,890  & $8.37\times 10^{-3}$ & $1.05\times 10^{-2}$ & $9.99\times 10^{-7}$ & $3.20\times 10^{-9}$  & $5.42\times 10^{-8}$ & $5.34\times 10^{-8}$ \\
			6.4 - 6.6 & 124,673  & $4.37\times 10^{-3}$ & $4.66\times 10^{-3}$ & $4.39\times 10^{-7}$ & $2.10\times 10^{-9}$  & $1.43\times 10^{-8}$ & $2.14\times 10^{-8}$ \\
			6.6 - 6.8 & 52,619   & $1.84\times 10^{-3}$ & $1.97\times 10^{-3}$ & $1.84\times 10^{-7}$ & $1.35\times 10^{-9}$  & $5.73\times 10^{-9}$ & $1.18\times 10^{-8}$ \\
			6.8 - 7.0 & 19,016   & $6.67\times 10^{-4}$ & $7.66\times 10^{-4}$ & $7.15\times 10^{-8}$ & $7.62\times 10^{-10}$ & $2.19\times 10^{-9}$ & $4.89\times 10^{-9}$ \\
			\hline
		\end{tabular}
	}
	\end{table*}

	\begin{table*}[htbp!]
	\caption{Information related to all-particle cosmic ray energy spectrum using two stations events. QGSJetII-04 is the hadronic interaction model assumption. Refer to Tab.~\ref{cr_flux_table} for detail description of each column. \label{cr_flux_table_qgsjet}}
	\resizebox{\textwidth}{!}{%
		\renewcommand{\arraystretch}{1}
		\begin{tabular}{lrcccccc}
			\hline
			$\log_{10}(E/{\rm GeV})$ & $N_{\rm events}$& Rate & Unfolded Rate& Flux           & Stat. Err  & Sys Low & Sys High \\
			& & [Hz] & [Hz] & \multicolumn{4}{c}{[$\rm m^{-2}s^{-1}sr^{-1}$]} \\
			\hline
			5.4 - 5.6 & 3,476,123& $1.22\times 10^{-1}$ & $1.15\times 10^{-1}$ & $2.11\times 10^{-5}$ & $1.34\times 10^{-8}$  & $3.58\times 10^{-6}$ & $2.06\times 10^{-6}$ \\
			5.6 - 5.8 & 2,731,596& $9.57\times 10^{-2}$ & $7.71\times 10^{-2}$ & $9.06\times 10^{-6}$ & $7.32\times 10^{-9}$  & $1.17\times 10^{-6}$ & $8.73\times 10^{-7}$ \\
			5.8 - 6.0 & 1,243,001& $4.35\times 10^{-2}$ & $4.54\times 10^{-2}$ & $4.48\times 10^{-6}$ & $5.35\times 10^{-9}$  & $6.46\times 10^{-7}$ & $3.90\times 10^{-7}$ \\
			6.0 - 6.2 & 484,928  & $1.70\times 10^{-2}$ & $2.33\times 10^{-2}$ & $2.18\times 10^{-6}$ & $4.22\times 10^{-9}$  & $3.43\times 10^{-7}$ & $1.96\times 10^{-7}$ \\
			6.2 - 6.4 & 269,906  & $9.45\times 10^{-3}$ & $1.05\times 10^{-2}$ & $9.62\times 10^{-7}$ & $2.37\times 10^{-9}$  & $1.23\times 10^{-7}$ & $7.76\times 10^{-8}$ \\
			6.4 - 6.6 & 107,815  & $3.78\times 10^{-3}$ & $4.10\times 10^{-3}$ & $3.74\times 10^{-7}$ & $1.25\times 10^{-9}$  & $3.46\times 10^{-8}$ & $2.97\times 10^{-8}$ \\
			6.6 - 6.8 & 48,760   & $1.71\times 10^{-3}$ & $1.69\times 10^{-3}$ & $1.53\times 10^{-7}$ & $8.37\times 10^{-10}$  & $1.21\times 10^{-9}$ & $1.41\times 10^{-8}$ \\
			6.8 - 7.0 & 18,932   & $6.63\times 10^{-4}$ & $6.87\times 10^{-4}$ & $6.23\times 10^{-8}$ & $5.69\times 10^{-10}$ & $4.47\times 10^{-9}$ & $4.47\times 10^{-9}$ \\
			\hline
		\end{tabular}
	}
	\end{table*}

\clearpage

\section{Experimental Data and Comparison with Sibyll2.1 Simulation} \label{appendix_data_mc_comparison}
	%------------------------
 	\begin{figure*}[ht!]
	\centering
	\includegraphics[height=6.8cm]{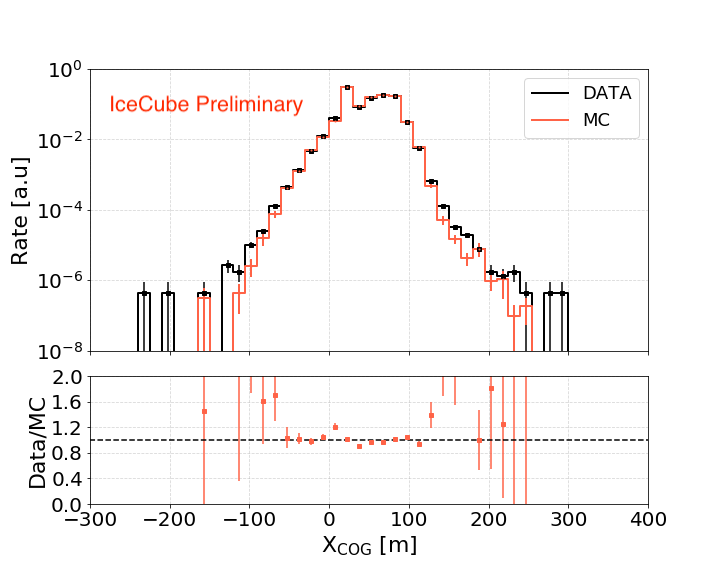}
	\includegraphics[height=6.8cm]{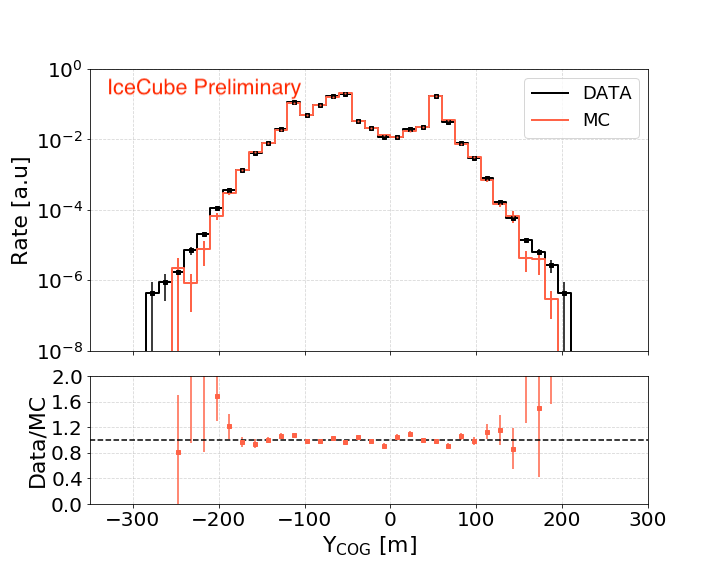}
	\caption{Histograms of the shower center of gravity from experimental data and simulation. The left plot is the x-coordinate and the right plot is the y-coordinate of shower cores. Peaks seen in both histograms are due to a larger number of tanks around that x or y coordinate. Refer to Fig.~\ref{it_geometry} for positions of all tanks.}
	\end{figure*}
	
    %------------------------
    \begin{figure*}[ht!]
	\centering
	\includegraphics[height=6.8cm]{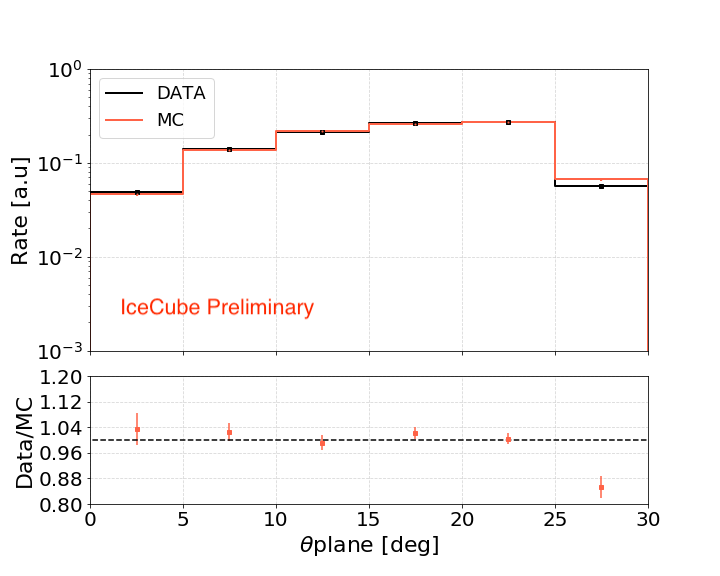}
	\includegraphics[height=6.8cm]{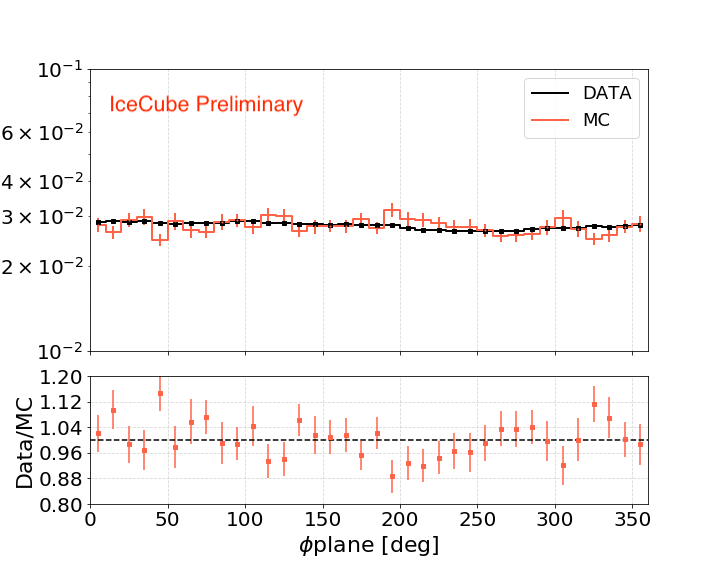}
	\caption{Histograms of zenith angle (left) and azimuth angle (right) calculated assuming plane shower front.}
	\end{figure*}
	
    %------------------------
    \begin{figure*}[ht!]
	\centering
	\includegraphics[height=6.8cm]{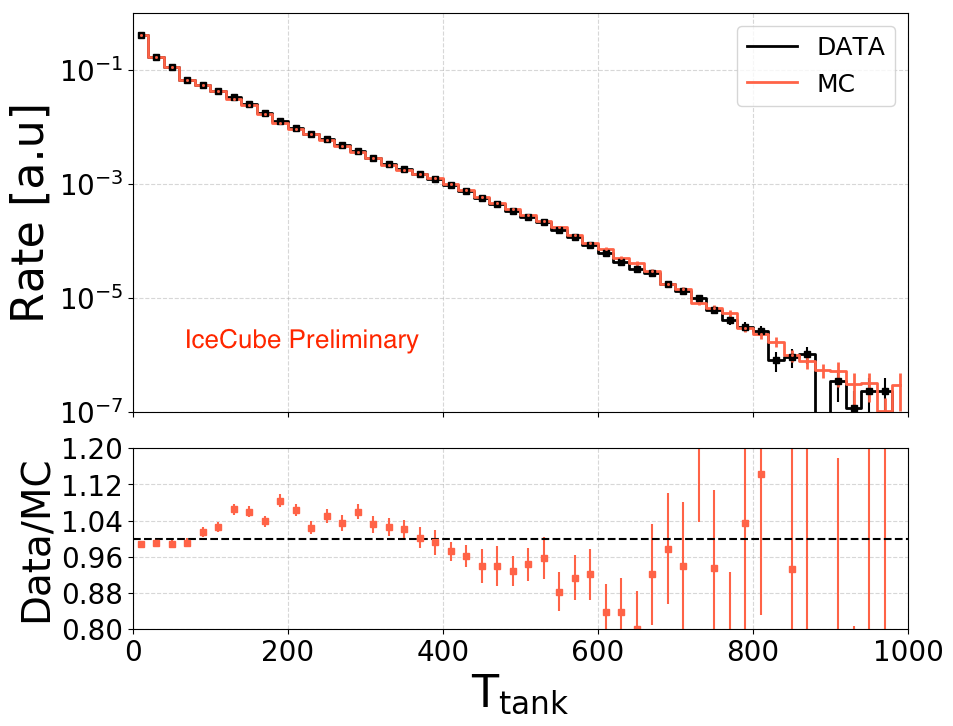}
	\includegraphics[height=6.8cm]{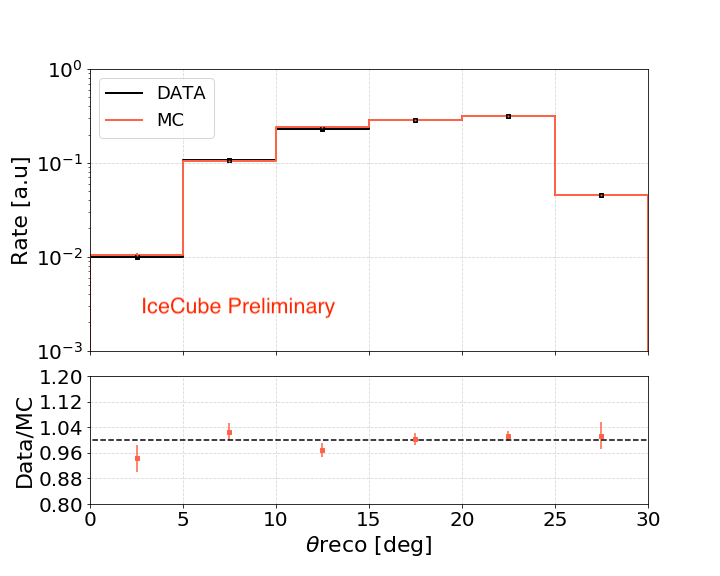}
	\caption{Left: Histograms of time difference of hits on each tank with respect to the first hit. Time of hits on each tank of an event is listed and sorted in ascending order. The time difference is with respect to the first hit. Time on hit tanks has high feature importance while reconstructing zenith angle. Right: Histograms of reconstructed zenith angle for experimental data and simulation. Cosine of reconstructed zenith angle is the third most important feature while reconstructing energy.}
	\end{figure*}
	
	%------------------------
    \begin{figure*}[ht!]
	\centering
	\includegraphics[height=6.8cm]{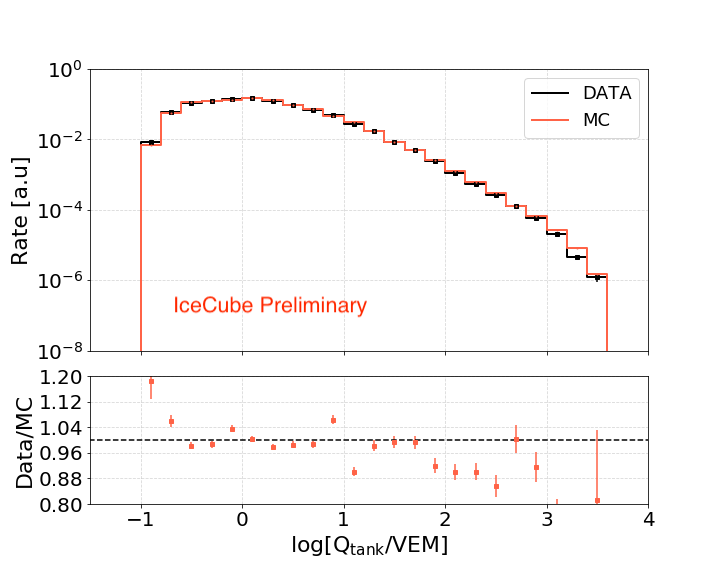}
	\includegraphics[height=6.8cm]{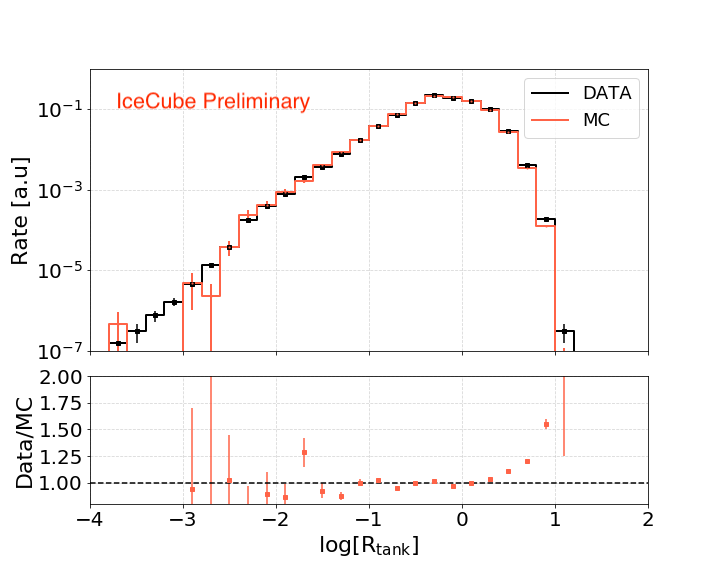}
	\caption{Left: Histograms of charge deposited on hit tanks. Charge on tanks has a high feature importance while reconstructing shower energy. Charge less than 0.16 VEM on a tank is considered due to background noise. Right: Histograms of the distance of hit tanks from the reconstructed shower core. The distance is divided by a reference distance of \unit[60]{m}. The list of distance of hit tanks from the core has high feature importance while reconstructing shower energy.}
	\end{figure*}
    %------------------------
    \begin{figure*}[ht!]
	\centering
	\includegraphics[height=6.8cm]{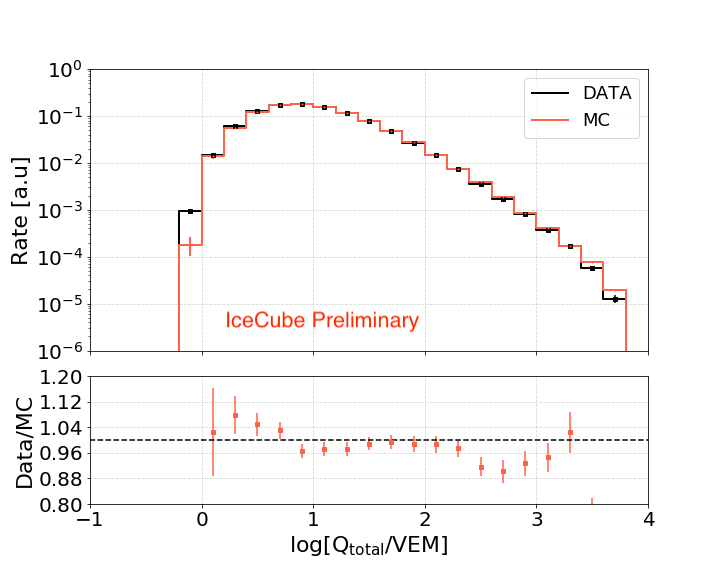}
	\caption{Histograms of the total amount of charge deposited in all stations. It has comparatively small feature importance while predicting energy.}
	\end{figure*}

\end{document}